\documentclass[a4paper,11pt]{article}
\usepackage{contribution}

\newcommand{\weblink}[2][]{%
    \ifthenelse{\equal{#1}{}}%
    {\textnormal{\url{#2}}}%
    {\textnormal{\href{#2}{#1}}}%
}

\def\beq{\begin{equation}}
\def\eeq#1{\label{#1}\end{equation}}
\def\eeqn{\end{equation}}

\def\beqa{\begin{eqnarray}}
\def\eeqa#1{\label{#1}\end{eqnarray}}
\def\eeqan{\end{eqnarray}}

\let\bar=\overbar

\def\Dslash{\not{\hbox{\kern-4pt $D$}}}
\def\dslash{\not{\hbox{\kern-2pt $\del$}}}

\def\msb{{\bar{\ssstyle M \kern -1pt S}}}

\newcommand{\contribution}[7][]{%
  \clearpage
  \thispagestyle{plain}
  \ifthenelse{\equal{#1}{}}
  {\hypersetup{pdftitle={#2}}}
  {\hypersetup{pdftitle={#1}}}
  \hypersetup{pdfauthor={{#3} {#4}}}
  {\centering\normalfont\LARGE\bfseries\sffamily #2 \par\nobreak}
  \lhead{}
  \chead{%
    \textit{\footnotesize XIV International Conference on Hadron Spectroscopy
      (\weblink[\textit{hadron2011}]{http://www.hadron2011.de}), 13-17 June 2011, Munich, Germany}%
  }
  \rhead{}
  \bigskip
  \begin{center}
    {#3} {#4}\ifthenelse{\equal{#6}{}}{}{\footnote{\weblink[#6]{mailto:#6}}}
    \ifthenelse{\equal{#7}{}}{}{#7} \\
    \textit{#5}
  \end{center}
  \bigskip
}

\renewcommand{\abstract}[1]{%
  \begin{center}
    \begin{minipage}{0.85\textwidth}
      \begin{footnotesize}
        #1
      \end{footnotesize}
    \end{minipage}
  \end{center}
  \bigskip
}

\begin{document}
{\makeatletter\@ifundefined{c@affiliation}
{\newcounter{affiliation}}{}\makeatother
\newcommand{\affiliation}[2][]{\setcounter{affiliation}{#2}
\ensuremath{{^{\alph{affiliation}}}\text{#1}}}
\contribution[Pion Elastic Form Factor from Local-Duality QCD Sum
Rule]{Pion Elastic Form Factor in a Rather Broad Range of Momentum
Transfers from Local-Duality QCD Sum Rule}
{Dmitri}{Melikhov}{\affiliation[HEPHY, Austrian Academy of
Sciences, Nikolsdorfergasse 18, A-1050 Vienna, Austria]{1}\\
\affiliation[Faculty of Physics, University of Vienna,
Boltzmanngasse 5, A-1090 Vienna, Austria]{2}\\\affiliation[SINP,
Moscow State University, 119991 Moscow, Russia]{3}}
{dmitri\_melikhov@gmx.de}
{\!\!$^,\affiliation{1}^,\affiliation{2}^,\affiliation{3}$, Irina
Balakireva\!\affiliation{3}, and Wolfgang Lucha\!\affiliation{1}}

\vspace{-.655ex}\abstract{Revisiting the relevance of local
duality for the pion elastic form factor gives rise to optimism.}

Recently, several analyses of the pion form factor $F_\pi(Q^2)$ at
momentum transfer $Q^2$ around $Q^2\approx4-50$ GeV$^2$ have
appeared \cite{recent} which claim that $F_\pi(Q^2)$ remains much
larger than the pQCD result even at $Q^2\approx 50$ GeV$^2$ (see
Fig.~\ref{Plot:1}a). These studies obtain a much larger
$F_\pi(Q^2)$ than our result from the local-duality sum rule
\cite{braguta}. They imply that the LD limit is strongly violated
even at rather large $Q^2$. QCD sum rules utilizing nonlocal
condensates \cite{bakulev} arrive at more moderate claims but also
observe a large local-duality violation at $Q^2=10-20$ GeV$^2$. A
careful inspection reveals that all these analyses involve
explicit or implicit assumptions. Consequently, in a recent study
\cite{irina} we scrutinized the LD model and its accuracy, by
taking advantage of the case of quantum mechanics: there hadronic
features, such as form factors, may be found independently of the
sum-rule method by solving the Schr\"odinger equation.

\begin{figure}[h!]\begin{center}\begin{tabular}{cc}
\includegraphics[width=7cm]{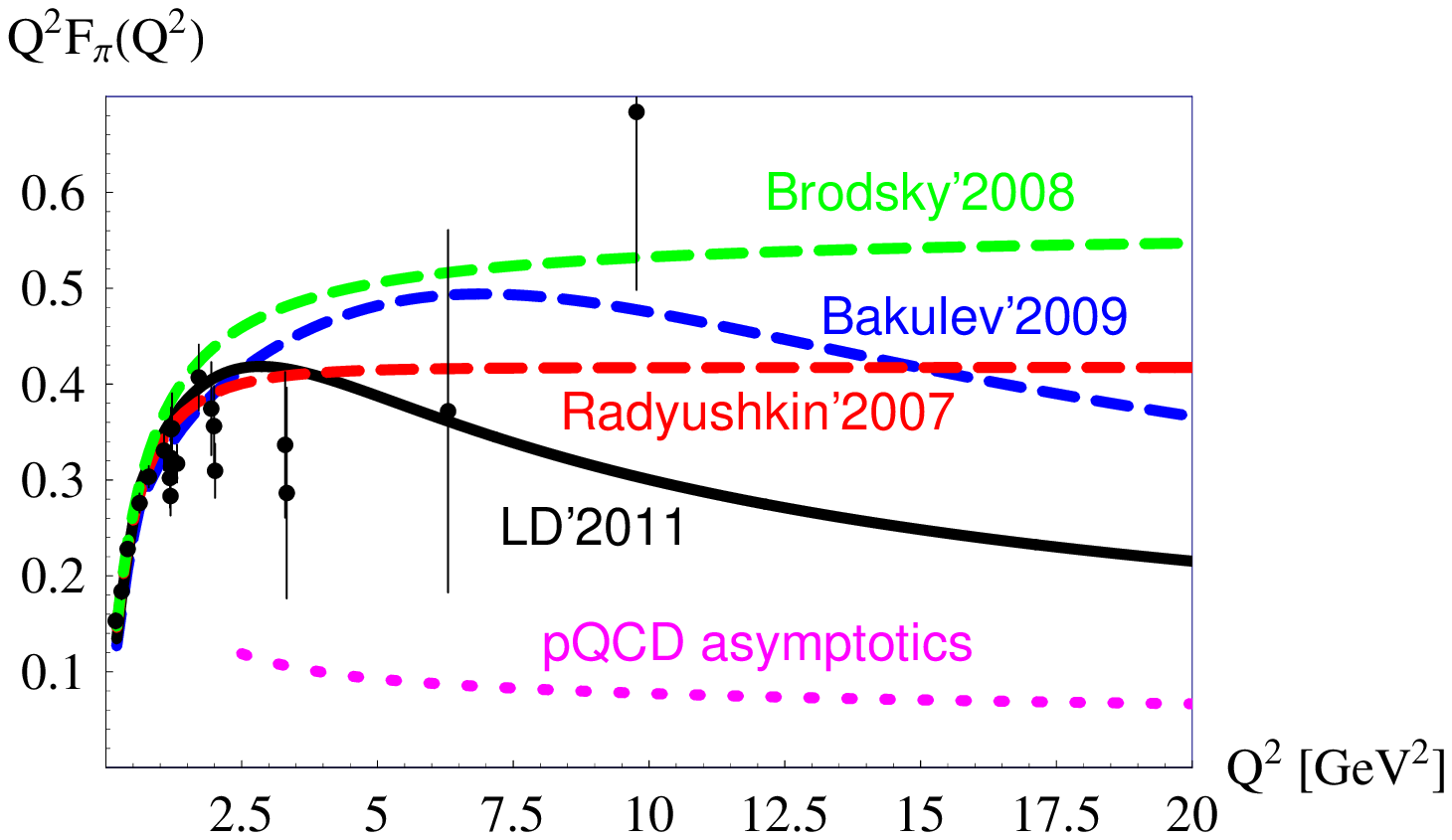}&
\includegraphics[width=7cm]{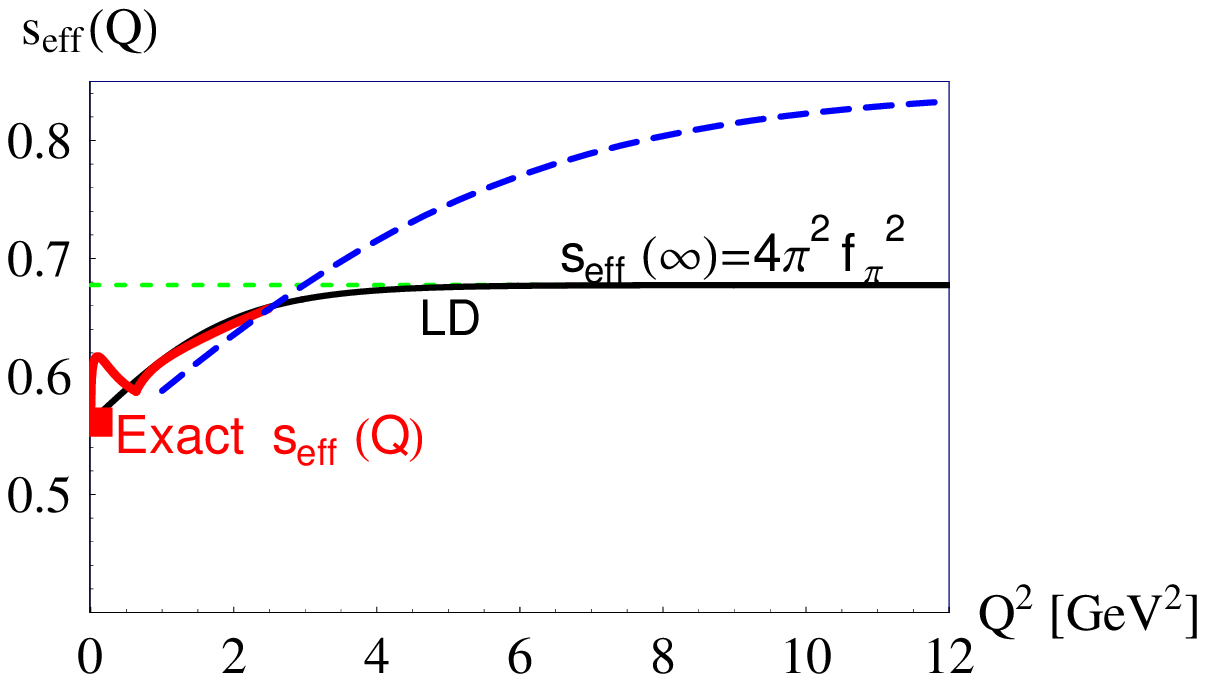}\\(a)&(b)
\end{tabular}\caption{\label{Plot:1}Dependence of both pion elastic
form factor $F_\pi(Q^2)$ (a) and effective continuum threshold
$s_{\rm eff}(Q)$ (b) on the momentum transfer $Q.$ In (b) the red
line is the exact threshold $s_{\rm eff}(Q)$ as reconstructed from
experimental $F_\pi(Q^2\le2.5\;\mbox{GeV}^2)$ data \cite{data},
the blue refers~to a sum-rule study using nonlocal condensates
\cite{bakulev}, the black is our LD model (\ref{seff_LD}) for
$s_{\rm eff}(Q)$.}\end{center}
\end{figure}

\section{Local-Duality Sum Rules}{\it Local-duality (LD) sum rules\/}
\cite{ld} are nothing but dispersive $N$-point sum rules in the
limit of infinitely large Borel-mass parameter $\widetilde M,$
that is, for $\tau\equiv1/\widetilde M^2\to0.$ In this limit, all
power corrections vanish. The {\it assumption\/} of {\it
quark--hadron duality\/} claims that, above some effective
continuum threshold $s_{\rm eff}$, the contributions of excited
and continuum states at~hadron level are dual to the high-energy
region of the perturbative diagrams arising from QCD. Under this
assumption, in the chiral limit the LD sum rules of interest for
the present analysis~read
\begin{equation}\label{SR_P5}
f^2_\pi=\int\limits_0^{\bar s_{\rm eff}}ds\,\rho_{\rm pert}(s)\
,\qquad F_\pi(Q^2)\,f^2_\pi=\int\limits_0^{s_{\rm
eff}(Q)}\,\int\limits_0^{s_{\rm eff}(Q)}ds_1\,ds_2\,\Delta_{\rm
pert}(s_1,s_2,Q)\ .\end{equation}The spectral densities $\rho_{\rm
pert}(s)$ and $\Delta_{\rm pert}(s_1,s_2,Q)$ are given by QCD
perturbation theory \cite{m}. All details of nonperturbative
dynamics are encoded in the effective continuum thresholds $\bar
s_{\rm eff}$ and $s_{\rm eff}(Q).$ Fixing these, pion decay
constant $f_\pi$ and form factor $F_\pi(Q^2)$ can be derived.

One should be aware that any effective continuum threshold is
different from the physical continuum threshold: the latter is a
constant determined by the masses of the lowest-lying hadronic
excitations whereas the effective continuum threshold is just an
ingredient of the sum-rule method related to a specific
implementation of quark--hadron duality. Therefore, effective
thresholds are not constant but depend on the external kinematic
variables \cite{lms_sr1,lms_sr2}.

Taking into account the properties of the perturbative 2- and
3-point spectral functions, one may formulate an approximate {\it
LD model for the effective threshold} $s_{\rm eff}(Q)$: our model
is based on some smooth interpolation between the behaviour of
$s_{\rm eff}(Q)$ for $Q\to0$, determined by a Ward identity, and
for $Q\to\infty$, determined by factorization properties of
$\Delta_{\rm pert}(s_1,s_2,Q)$. Remembering the well-measured pion
elastic form factor in the region near $Q^2\approx2.5$ GeV$^2$, we
propose \cite{irina}, in terms of the strong coupling constant
$\alpha_s(Q^2),$ the simple parametrization
\begin{equation}\label{seff_LD}
s_{\rm eff}(Q)=\frac{4\pi^2\,f^2_\pi}{1+\alpha_s(0)/\pi}
\left[1+\tanh\!\left(\frac{Q^2}{Q_0^2}\right)\frac{\alpha_s(0)}{\pi}\right],
\qquad Q_0^2=2.02\;{\rm GeV}^2\ .\end{equation}For small $Q^2$,
following \cite{braguta} we assume a freezing of $\alpha_s(Q^2).$
Note that $s_{\rm eff}(Q)$ approaches its LD limit, viz., $s_{\rm
eff}=4\pi^2\,f_\pi^2$, already in the region $Q^2>4-5$ GeV$^2$
(Fig.~\ref{Plot:1}b). Accordingly, the only essential
nonperturbative input for the LD model (\ref{seff_LD}) is the pion
decay constant~$f_{\pi}$.

Figure \ref{Plot:1}a depicts the corresponding prediction for the
pion elastic form factor $F_\pi(Q^2).$ Our LD model provides a
perfect description of all the available experimental data in the
region $Q^2=1-2.5$ GeV$^2.$ For $Q^2\ge3-4$ GeV$^2,$ the LD model
reproduces well all the data except for the single point $Q^2=10$
GeV$^2$; there our prediction is off the actual experimental value
(which, in any case, is affected by a rather large error) by
roughly two standard deviations. Interestingly enough, in the
range $Q^2\ge3-4$ GeV$^2$ the LD model yields significantly~lower
predictions than the findings of the different theoretical
approaches presented in Refs.~\cite{recent,bakulev}.

A closer inspection of Fig.~\ref{Plot:1} easily reveals that it is
virtually impossible to construct models compatible with all
experimental findings in $Q^2=2.5-10$ GeV$^2$: those approaches
that hit the data at $Q^2=10$ GeV$^2$ overestimate the data points
of better quality at $Q^2\approx2-4$~GeV$^2$.

By construction, the LD model (\ref{seff_LD}) is but an
approximate model which involves too few free parameters to be
able to take into account some subtle details of the confinement
dynamics. Nevertheless, we would like to estimate the
uncertainties of hadron-parameter predictions we might expect for
the momentum-transfer range $Q^2\ge3-4$ GeV$^2.$ The obvious place
to study this and to get an idea of the order of magnitude of the
errors is quantum mechanics: there solving Schr\"odinger's
equation numerically \cite{Lucha98} gives the exact bound-state
features.

\section{Local-Duality Effective-Threshold Model in Quantum
Mechanics}The main ingredient that constrains the formulation of
our LD model (\ref{seff_LD}) is the factorization of hard form
factors. Consequently, this model may be tested in quantum
mechanics (QM) for potentials containing both Coulomb and
confining interactions. For definiteness, we \cite{irina} consider
a set of power-law confining potentials: $V_{\rm conf}(r)\propto
r^n,$ $n=2,1,\frac{1}{2}.$ We adopt model parameters suitable for
hadron physics and fix the strengths of all our $V_{\rm conf}(r)$
such that for each of them the Schr\"odinger equation yields the
{\em same value\/} $\psi(0)$ of the \mbox{configuration-space}
bound-state wave function $\psi(r)$ at the origin and hence the
same QM LD threshold model.

We identified an important {\em universal\/} behaviour, which does
not depend on the details of the confining interaction: the
accuracy of the LD model for both effective continuum threshold
and elastic form factor increases with $Q$ in the range $Q\ge2$
GeV. Accordingly, we may infer that, if in the region
$Q^2\approx4-8$ GeV$^2$ the LD setup provides a satisfactory
description of the experimental data, the accuracy of our
predictions will not be worse for larger values~of~$Q^2.$

\section{Summary, Conclusions, and Outlook}We investigated the
pion elastic form factor $F_\pi(Q^2)$ by means of an LD model,
which can be formulated in any theory where hard exclusive
amplitudes satisfy a factorization theorem (in essence, any theory
where the interactions behave Coulomb-like at small distances and
confining at large distances). Figure \ref{Plot:1} and our QM
studies lead us to our main conclusions:

{\bf 1.} For $Q^2\le 4$ GeV$^2$, our exact effective threshold
$s_{\rm eff}(Q)$ exhibits a rapid variation with $Q$. This
observation implies that the accuracy of the LD model for these
momentum transfers depends on subtle details of the confining
interactions and cannot be predicted in advance.

{\bf 2.} For $Q^2\ge 4$ GeV$^2$, irrespective of any details of
the underlying confining interactions, the {\em maximum\/}
deviations of the LD-model predictions from the exact elastic form
factor occur in the range $Q^2\approx 4-8$ GeV$^2$. For $Q^2$
beyond this interval, our LD model's accuracy increases very fast.
Our QM toy model with power-law potentials shows that, for
arbitrary confining interactions, our LD model entails rather
accurate numerical results for $Q^2\ge20-30$ GeV$^2.$

\newpage{\bf 3.} Very precise data [5] on $F_\pi(Q^2)$ indicate
that the LD limit $s_{\rm eff}(\infty)=4\pi^2\,f_\pi^2$ of the
effective threshold is reached already at comparatively low values
$Q^2=5-6$ GeV$^2$; therefore, large deviations from the LD limit
at $Q^2=20-50$ GeV$^2,$ as obtained in \cite{recent}, appear to us
unlikely. Moreover, we expect the pion form factor at $Q^2=10-20$
GeV$^2$ to be considerably lower than the prediction of an
approach based on a sum rule involving nonlocal condensates
\cite{bakulev}.

Our analysis is not meant to constitute a proof of but rather to
provide an argument for the {\em accuracy\/} of the LD model in
QCD and the expected behaviour of the pion elastic form factor at
large $Q^2$. Thus, the accurate measurement of $F_\pi$ in the
region $Q^2=4-10$ GeV$^2$ will have important implications for the
behaviour of $F_\pi$ at larger $Q^2,$ up to asymptotically~large
$Q^2.$

\vspace{.1ex}\noindent{\em Acknowledgments.} DM is supported by
the Austrian Science Fund (FWF), project
no.~P22843.}

\end{document}